%% file: StockNNetDiversification.tex
\documentclass[a4paper,onecolumn,12pt,epsf]{article}
\usepackage{epsfig}
\usepackage{epstopdf}
\usepackage{rotating}
\usepackage{amsmath}
\usepackage{longtable}
\usepackage{natbib}         % citation style
\begin{document}

\title{An Application of Correlation Clustering to Portfolio Diversification}
\author{Hannah Cheng Juan Zhan$^{1}$,   William Rea$^{1}$,  and Alethea Rea$^{2}$, \\
1. Department of Economics and Finance, University of Canterbury, \\
New Zealand \\
2. Data Analysis Australia, Perth, Australia }

\maketitle

\input Abstract.tex

%\maketitle

\input Intro.tex

\input Data.tex

%\input nnetdesc.tex

\input Methods.tex

\input FindingClusters.tex

\input MovingIndustries.tex

\input Example.tex

\input Discussion.tex

%\input Conclusion.tex

%\clearpage

\bibliography{StockNNetDiversification}

\clearpage

\appendix

\input 126Companynames.tex

%\input  stockcodes.tex

%\clearpage
\input Period1Clusters.tex

%\input P2Clusters.tex

%\input P3Clusters.tex

%\input P4Clusters.tex

%\input ExtraNNDiagrams.tex

%\newpage

%\input OtherNetworks.tex

%\bibliography{StockNNet}

\end{document}

%% file: Abstract.tex
\begin{abstract}
This paper presents a novel application of a
 clustering algorithm developed for constructing
a phylogenetic network to the correlation matrix for 126 stocks listed on
the Shanghai A Stock Market. We show that by visualizing the
correlation matrix  using a Neighbor-Net network and using the circular ordering 
produced during the construction of the network we can reduce the risk
of a diversified portfolio compared with random or industry group based
selection methods in times of market increase.
\end{abstract}

\begin{description}
\item[Keywords: ] Visualization, Neighbour-Nets,  Correlation Matrix,
Diversification, Stock Market
\item[JEL Codes: ] G11
\end{description}

%% file: Intro.tex
\section{Introduction}\label{sec:introduction}

Portfolio diversification is critical for risk management because
it aims to reduce
the variance in returns compared with a portfolio of a single stock or similarly
undiversified portfolio. 
The academic literature on diversification is vast,
stretching back at least as far as \cite{lowenfeld1909}. The modern
science of diversification is usually traced to \cite{markowitz1952}
which is expanded upon in great detail in \cite{markowitz1959}.

The literature
covers a wide range of approaches to portfolio diversification, such as;
the number of
stocks 
required to form a well diversified portfolio,
which has increased from eight stocks 
in the late 1960's \citep{Evans1968} to over 100 stocks in the late 2000's \citep{Domian2007},
what types of risks should be considered, \citep{Cont2001, 
Goyal2003, Bali2005}, factors intrinsic to each
stock \citep{Fama1992, Fama1993},
the age of the investor, \citep{Benzoni2007}, and
whether international diversification is beneficial, \citep{Jorion1985,Bai2010}, among others.

Despite the recommendation of authorities like \cite{Domian2007},
 \cite{Barber2008} reported that in a large sample of 
American private investors
the average portfolio size of individual stocks was only 4.3. 
While comparable data does not appear to be available for
private Chinese investors, it seems unlikely that they hold
substantially larger portfolios.

The mean returns and variances of the individual contributing stocks are insufficient
for making an informed decision on selecting a suite of stocks because 
selecting a portfolio requires an understanding of the correlations between each of the stocks
being considered for the portfolio. The number
of correlations between stocks rises in proportion to the square of
the number of stocks meaning that for all but the smallest of stock
markets 
the very large number of correlations is beyond
the human ability to comprehend them.  \cite{Rea2014} 
presented a method
to visualise the correlation matrix using
neighbor-Net networks \citep{Bryant2004}, yielding insights into the relationships between 
the stocks.

Neighbor-Net networks are widely used in other fields, for example
document source critical analysis
\citep{Tehrani2013}, understanding the cultural geography of folktales 
\citep{Ross2013}, understanding human history through language (\cite{Grey2010, Heggarty2010, Knooihuizen2012} among others), understanding human history through housing traditions \citep{Jordan2010} and to study the evolution of the skateboard deck \citep{Prentiss2011}. Its main application area is biology where neighbor-Net networks have appeared in hundreds of refereed papers. Recently these networks have been used to assist in understanding cancer (see \cite{Schwarz2015} for an example), investigate the evolution of high mountain buttercups \citep{Emadzade2015} and study mosquito borne viruses \citep{Bergqvist2015}. 

Traditional investing wisdom has suggested that 
investors should select investment
opportunities from a range of industries because
returns within an industry would be
more highly correlated than those between industries. While that may hold true,
there are some instances (such as companies with operations in several industries)
in which a stock exchange industry classification alone is insufficient.
Furthermore with some authors (including \cite{Domian2007}) recommending
over 100 investments, the number of investments may exceed the number of industries
meaning there is a need to select a diverse range of
stocks even within industries. 

Another key aspect of stock correlation is the potential change in the correlations with a 
significant change in market conditions (say comparing times of general market increase with 
recession and post-recession periods). %maybe remove depending on focus

%Here we need more about portfolio diversification

In this paper we explore investment opportunities in China using data from the Shanghai
Stock Exchange. We compare the correlation structure reported in four periods (a period of market calm 2005/2006, a boom 
period of 2006/2007, market decline (2008), and a
post crash period 2009/2010). 

Our primary motivation is to investigate four portfolio selection
strategies.  The four strategies are;
\begin{enumerate}
\item picking stocks at random;
\item forming portfolios by picking stocks from different industry groups;
\item forming portfolios by picking stocks from different correlation clusters; and
\item forming portfolios
by picking stocks from industry groups within correlation clusters.  
\end{enumerate}
Our results show that
knowledge of correlations clusters can  reduce
the portfolio risk.

The outline of this paper is as follows; Section (\ref{sec:data})
discusses the data, Section (\ref{sec:methods}) discusses the methods
used in this paper, Section (\ref{sec:clusters}) discusses
identifying
the correlation clusters, Section (\ref{sec:movements})
discusses the movement of stocks in the neighbor-Net splits
graphs between study periods, Section (\ref{sec:example}) applies
the results of the previous two sections to the
problem of forming a diversified portfolio of stocks,
and Section (\ref{sec:discussion}) contains
the discussion and our conclusions.

%% file: Data.tex
\section{Data}\label{sec:data}

The data used in this study was downloaded from Datastream. 
We obtained daily closing prices and dividend
data for 126 stocks from the Shanghai A Index. 
The data listed the stock name, a six digit 
identification number, and assigned the stock to one of five industry
groups. These groups were (1) Energy (12 stocks), (2) Finance
 (17 stocks), (3) Health Care (18 stocks),
(4) Industrial (33 stocks), and
(5) Materials (36 stocks). 
To make the identification of the stocks
and their exchange-assigned industry groups simpler we
generated four letter stock codes and to this code appended
a single letter indicating its industry group. A
list of these can be found in Table (\ref{tab:stockcodes})
in \ref{sec:stockcodes}.
To estimate stock return correlations we
calculated weekly returns from the daily price and dividend
data. To obtain the period returns we calculated the total
return for each period and treated the
dividends as being reinvested into the stock that issued them.

\begin{figure}[ht]
  \centering
  \includegraphics[width=12cm]{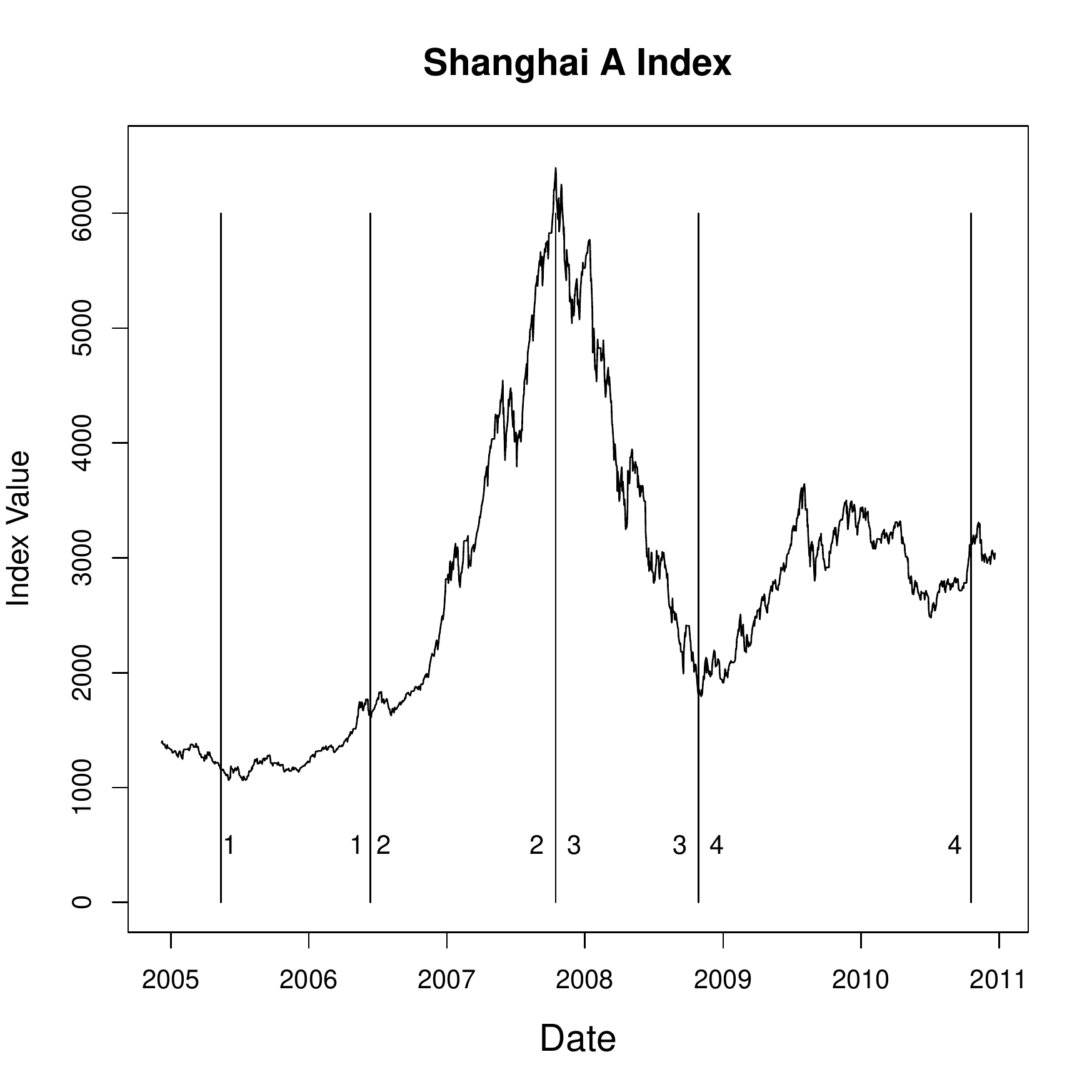}
  \caption{A plot of the Shanghai Stock Exchange A Index with the
boundaries of the four 
study periods marked. The dates are 13-May-2005,
13-June-2006,  16-Oct-2007,  28-Oct-2008, and  19-Oct-2010 respectively.
}
  \label{fig:ShnaghaiIndex}
\end{figure}

A graph of the index and the boundaries of our study periods
can be found in Figure (\ref{fig:ShnaghaiIndex}).
We defined the study periods so that they represented as 
different market conditons as we could make them, though
it could be argued that our study periods one and four are
similar.

Study period one was 13 May 2005 until 13 June 2006 and was a
period in which the market underwent a slow rise. Study period two was
13 June 2006 until 16 October 2007 and is a considered
a boom or market bubble period. Study period three was 16 October 2007 
until 28 October 2008 representing a sharp decline or crash.
The final study period was from 29 October 2008 until
19 October 2010 was a time of initial market recovery and then
a largely flat returns.

With four study periods, for the portfolio selection methods
which require a model building, or estimation, period we
can form models in periods one through three and use
the periods two through four for out-of-sample testing.
Such extremely different market conditions represents a
very severe test of portfolio diversification strategies,
especially forming portfolios based on period two and
testing them against period three data.

%-In period three, three of the stocks were not traded at all and as such they were 
%removed from the analysis. These were GFSC, XAAE and RSNM. 
%
% The spreadsheet shows  GFSC, XAAE and RSNM as being actively traded.

%-How were the returns calculated noting that they should include dividends and the like
% Hannah didn't actually do it like we say and so we should recalculate
% the per period returns.

%-Would be good to mention a little bit about the five industries and their relative sizes. 

%% file: Methods.tex
\section{Methods}\label{sec:methods}

\subsection{Neighbor-Net Splits Graphs}
A typical stock market correlation matrix for $n$ stocks
 is of full rank which means 
that it can only be represented fully in an $(n-1)$-dimensional
space. Some basic statistics on the correlations are
presented in Table (\ref{tab:basicstats}).
In visualization, the high dimensional data space is collapsed
to a much lower dimensional space so that the data can be
represented on 2-dimensional surface such as a page or computer
screen.

We
need to convert the numerical values in the correlation matrix to
a measure which can be construed to be a distance. In the
literature the most common way to do the conversion is
by using the so-called ultra-metric, 
\begin{equation}
d_{ij}=\sqrt{2(1-\rho_{ij})} \label{eqn:sqrt1mc} 
\end{equation}
where $d_{ij}$ is the estimated distance and
$\rho_{ij}$ is the estimated correlation between stocks $i$ and $j$,
see \cite{Mantegna1999}
for details.

Using the conversion in Equation \eqref{eqn:sqrt1mc} we formatted
the converted correlation matrix and 
augmented it with the appropriate stock codes for reading into
the Neighbor-Net software, SplitsTree \citep{Huson2006}, available from
 \verb+http://www.splitstree.org+. Using the SplitsTree software
we generated the Neighbor-Nets splits graphs. Because the splits
graphs are intended to be used for visualization we defer the
discussion of the identification of correlation clusters and
their uses to Sections (\ref{sec:clusters}) and (\ref{sec:movements})
below.

\subsection{Simulated Portfolios}

Recently \cite{Lee2011} discussed so-called risk-based asset
allocation. In contrast to strategies which require both
expected risk and expected returns for each investment opportunity
as inputs to the portfolio selection process, 
risk-based
allocation considers only expected risk. The four methods 
of portfolio selection we
present below can be considered to be risk-based allocation methods.
This probably reflects private investor behaviour in that often they
have nothing more than broker buy, hold, or sell recommendations
to assess likely returns.

The four portfolio methods were compared using simulations. For each of 1,000 iterations a portfolio was sampled based on the rules governing the 
portfolio type. We recorded the mean and standard deviation of the returns 
for the 1,000 portfolios.

As mentioned in the introduction the primary motivation is to investigate four portfolio strategies. These are:
\begin{enumerate}
\item Selecting stocks at random;
\item Selecting stocks based on industry groupings;
\item Selecting stocks based on correlation clusters; and
\item Selecting stocks based on industry groups within correlation clusters.
\end{enumerate}
We describe each of these in turn.

\begin{description}
\item[Random Selection: ]
The stocks were selected at random using a uniform distribution without 
replacement. In other words each stock was given equal chance of 
being selected according but with no stock being selected twice within a 
single portfolio. 

\item[By Industry Groups: ]
There were five industry groups. If the portfolio size was five or 
less, the industries were chosen at random using a 
uniform distribution without replacement.
From each of the selected industry groups one stock was selected. 
If the desired portfolio size was more than five then each group had at 
least $s$ stocks selected, where $s$ is the quotient of the portfolio 
size divided by five. Some (the remainder of the portfolio size divided by five) industry groups will have $s+1$ stocks selected and the industry groups this applied to were chosen using a uniform distribution without
replacement. Within each industry group stocks were selected using
a uniform distribution, again without replacement.
%Because there were only a few stocks selected from each industry group, portfolios with a stock selected twice remained in the sample. 
%% if we do want to check for stock double up I can add this to the code

\item[By Correlation Clusters: ]
%As described above, the Neighbor-Net diagrams are constructed from a distance matrix and represent the correlation structure. The LASSO algorithm for determining which splits are essential to represent the distances, returns an ordering of the splits, from the first to be included to the last. This ordering excludes any splits which were added in an iteration but later removed. This ordering provides a basis for determining eligible clusters.

%Many of the splits are trivial splits, those separating one taxa (stock) from the rest. Some splits separate small clusters of stocks from the remainder. These small clusters are likely to be too small to contain meaningful correlation clusters. Therefore an eligible correlation cluster is defined as a cluster larger than the cutoff value. In this paper we used 25\% of the total number of stocks as the cutoff (rounded down). 

%Each eligible cluster is defined by a split, or bipartition of the stocks. In this paper only the smaller bipartition was considered. The clusters can overlap and often have many stocks in common. The eligible clusters are ordered, essential for investigating the effect of varying portfolio size.

%Description above is for statistical clusters, we have used by eye

The correlation clusters were determined by examining the Neighbor-Net network for the relevant periods (period one, two and three). Each stock was assigned to exactly one cluster and each cluster can be defined by a single split (or bipartition) of the circular ordering of the Neighbor-Net of the relevant period. The clusters determined in periods one, two and three were used to generate the portfolios for out-of-sample
testing in periods two, three and four respectively. Because the
goal of portfolio building is to reduce risk each cluster was paired
with another cluster which was considered most distant from it.
This method is discussed in detail below.

As with the industry groups, if there were fewer clusters than the desired portfolio size, cluster pairs were selected at random and a stock selected from within each correlation cluster pair. 
If the desired portfolio size was larger than the number of correlation cluster then we apply the method described above for the industry clusters.

As indicated above each cluster was paired with the one most distant
from it. Because we identified
an even number of clusters in period two, cluster one was paired
with cluster five, two with six and so on. In periods with an
odd number of clusters the pairing may not be so straight-forward.
For example, in period two (see Figure \ref{fig:ChinaP2ClCC})
we identified five clusters and
cluster one was paired with four, both clusters two and three
were paired with five, four was paired with one and five with two.

\begin{figure}[ht]
  \centering
 \includegraphics[width=12cm]{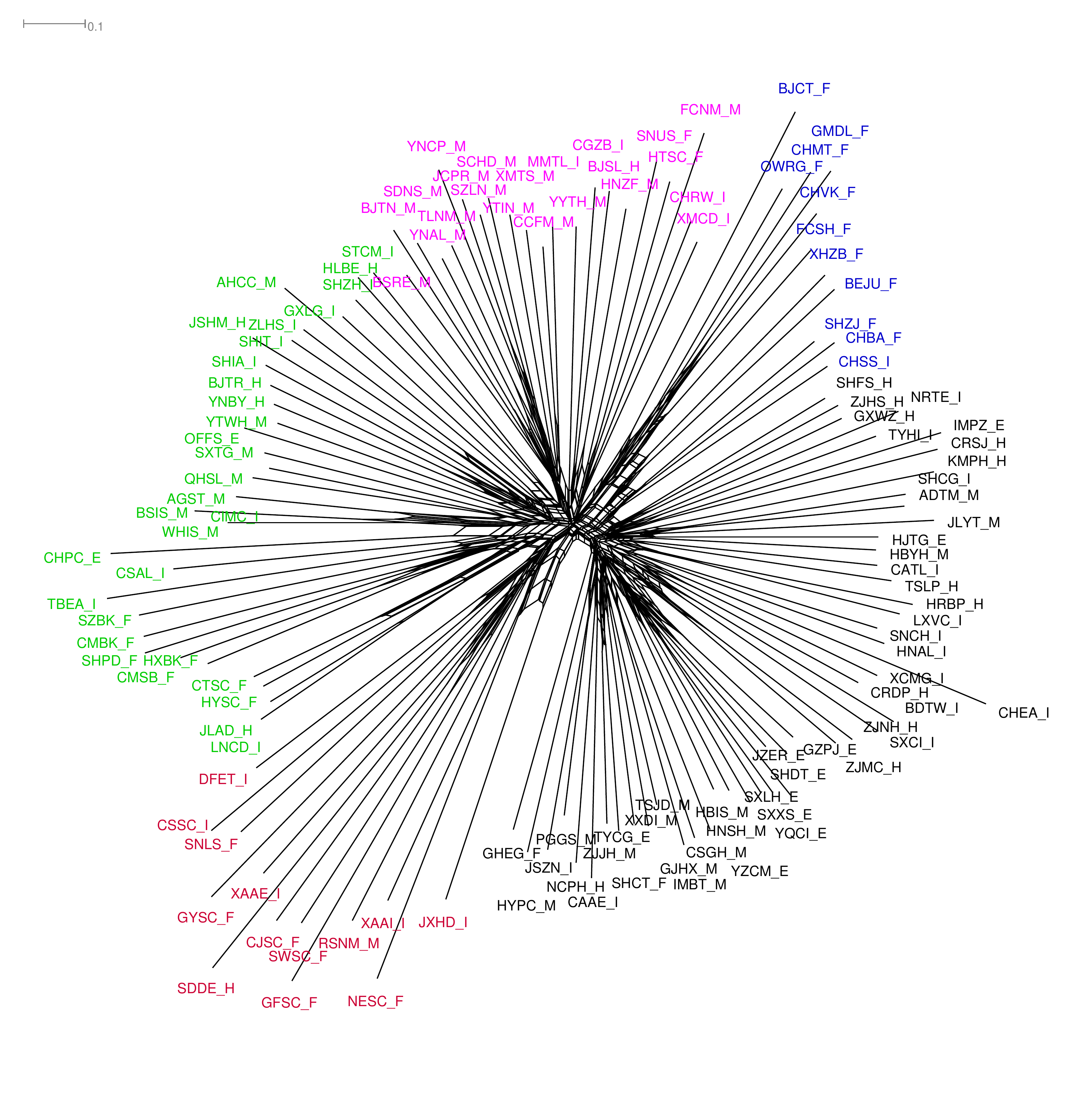}
  \caption{SplitsTree network for 126 stocks from the
Shanghai A Stock Exchange  for period two
using five trading day returns to estimate correlations
and hence distances with the stocks in cluster one colour
coded. The five correlation clusters each have different
colours. In the discussion the clusters are coded anti-clockwise
as follows;
Cluster 1 -- Black, Cluster 2 -- Blue, Cluster 3 -- Purple,
Cluster 4 -- Green,
Cluster 5 -- Red.}
  \label{fig:ChinaP2ClCC}
\end{figure}

\item[By Industry Group within Correlation Clusters: ]
The final method was selecting stocks  from industry groups within correlation clusters. Each stock within each  cluster has an associated industry group. Therefore each correlation cluster can be subdivided into up to five sub-clusters based on industry. 

As indicated above each cluster was paired with the one most distant
from it. Once a cluster was selected for inclusion, so was the paired cluster,
however this time we did not allow any of the paired stocks to be from the same
industry. This was the method used for determining the set of stocks for the fourth portfolio strategy. 
%If the group is below a minimum size it was discarded from the eligible set of stocks. The minimum size was set at 10 for this investigation.
%This wasn't how it was implemented

\end{description}

%% file: FindingClusters.tex
\section{Identifying Correlation Clusters}\label{sec:clusters}

\begin{figure}[ht]
  \centering
  \includegraphics[width=12cm]{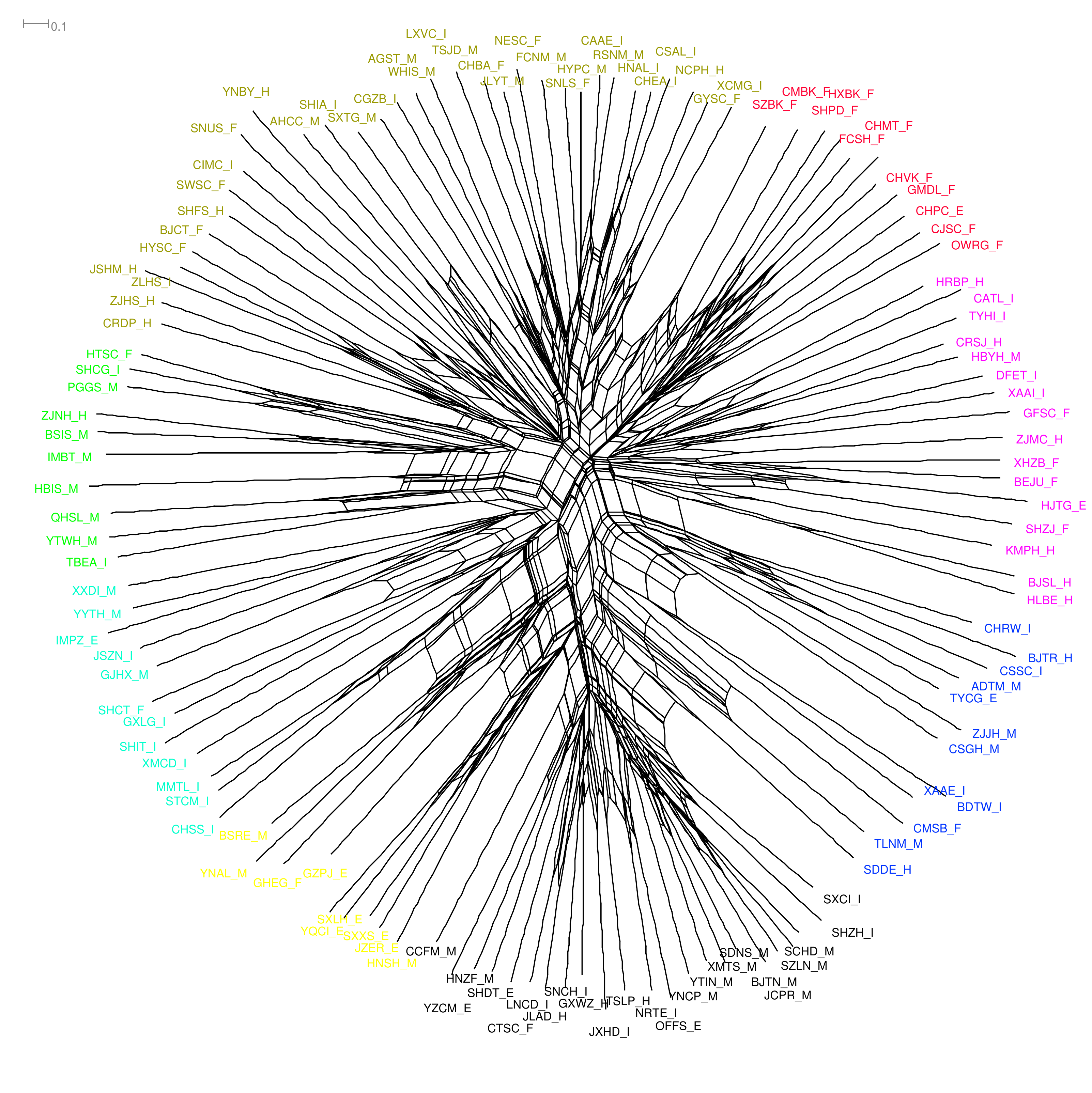}
  \caption{SplitsTree network for 126 stocks from the
Shanghai A Stock Exchange  for period one
using five trading day returns to estimate correlations 
and hence distances with the stocks in cluster one colour
coded. The eight correlation clusters each have different
colours. In the discussion the clusters are coded as follows;
Cluster 1 -- Black, Cluster 2 -- Blue, Cluster 3 -- Purple, 
Cluster 4 -- Red, Cluster 5 - Khaki, Cluster 6 -- Green,
Cluster 7 -- Aqua, Cluster 8 -- Yellow.}
  \label{fig:ChonaP1CC}
\end{figure}

As \cite{Bryant2004} point out ``the splits graphs generated by Neighbor-Net are
{\em always} planar, an important advantage over other network
methods when it comes to visualization'' (emphasis original). Thus
one method of identifying a group of stocks clustered by
correlation is to examine the splits graph for the stocks (see,
for example, Figure \ref{fig:ChonaP1CC}) and look for
natural breaks in the structure of the network. 

The neighbor-Net
splits graph is a type of map. All readers of a topographic map
read the map in the same way. The information they extract
depends on their needs. One person may read a map to extract information
about mountain ranges, another for information on river catchments,
and still another on the distribution of human settlements. But in
all cases all map readers agree which features are mountains, which
are rivers and which are towns and cities, no confusion arises because
the map is read visually.

Because this
is a visual approach, the information extracted
from reading a neighbor-Net
splits graph depends on
the researcher or financial analyst
balancing whatever competing requirements they may have. 
Here we know that in the simulations to follow
the sizes of the portfolios we
will generate will be two, four, eight or 16 stocks. Consequently
we do not need large numbers of clusters
and we would like them to have
a sufficiently large number of stocks that when selecting stocks
at random from within the cluster that there are a sufficiently
large number of combinations available to make the simulations
meaningful. 
These requirements guide us when identifying clusters in the
neighbor-Net splits graphs. The numbers of clusters and cluster
membership is determined visually and it is important not to confuse
visual with subjective. 
For period one we chose eight clusters, which was
the maximum number of clusters in any period. The smallest cluster had
nine stocks giving $\binom{9}{2}=36$ distinct ways of choosing
two stocks from this cluster in the 16 stock portfolio simulation.
 
Figure (\ref{fig:ChonaP1CC}) shows the clusters we identified for
period one.  The stocks in each cluster are listed in 
\ref{sub:period1}. 
Cluster one is at the bottom in black
and the clusters are sequentially numbered moving counter-clockwise around
the splits graph. Cluster one can be recognised by the small,
but clear, gaps in the network
structure between it and clusters two and eight. Similar
small gaps can be seen between the other clusters. 

This grouping
of eight clusters is not the only division of the stocks into clusters
which could have been made. If the
researcher or financial analyst had other requirements some of
the clusters could be further subdivided or combined. For example
if small clusters were acceptable then Cluster 2 could 
be further split into two clusters, as could Cluster 8. 
In both cases there is a clear gap in the network structure
where the split could be made. Conversely, if the number of
clusters desired was reduced then there are some reasonably clear
combinations which could be made. For example, if only two
clusters were required, then, perhaps, Clusters 1, 2, 7, and 8
could be combined to form one cluster while Clusters 3, 4, 5, and
6 would form the other.

%% file: MovingIndustries.tex
\section{Movements of Stocks in the Splits Graphs between Periods}
\label{sec:movements}

\begin{figure}[ht]
  \centering
  \includegraphics[width=12cm]{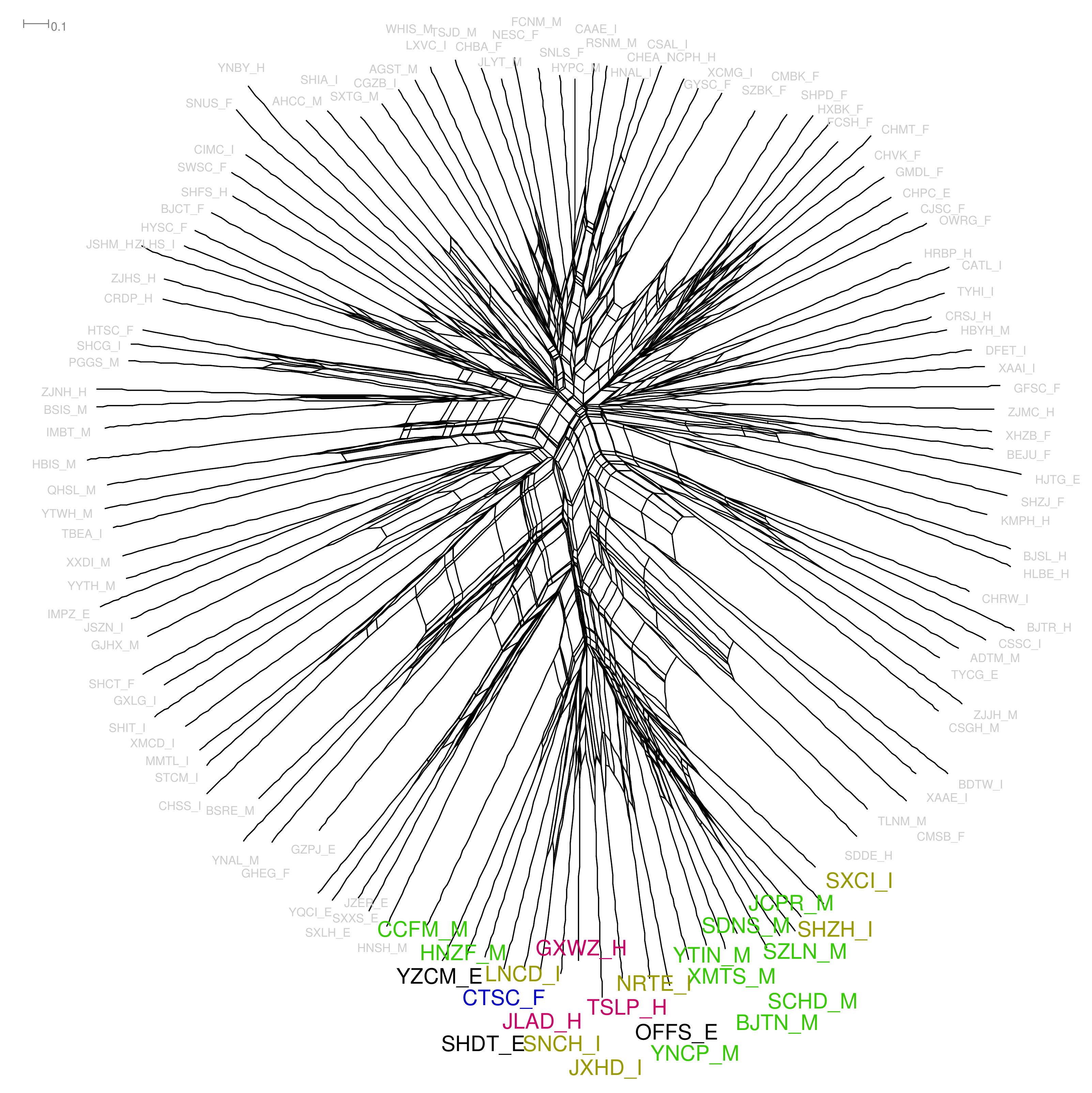}
  \caption{The SplitsTree network for the
Shanghai A Stock Exchange  for period one
with the stocks in cluster one colour
coded by industry group. The colours are Energy - Black, Finance -- Blue,
Health Care -- Red, Industrials -- Khaki, Materials -- Green.  }
  \label{fig:ChinaP1Cl1CC}
\end{figure}

\begin{figure}[ht]
  \centering
  \includegraphics[width=12cm]{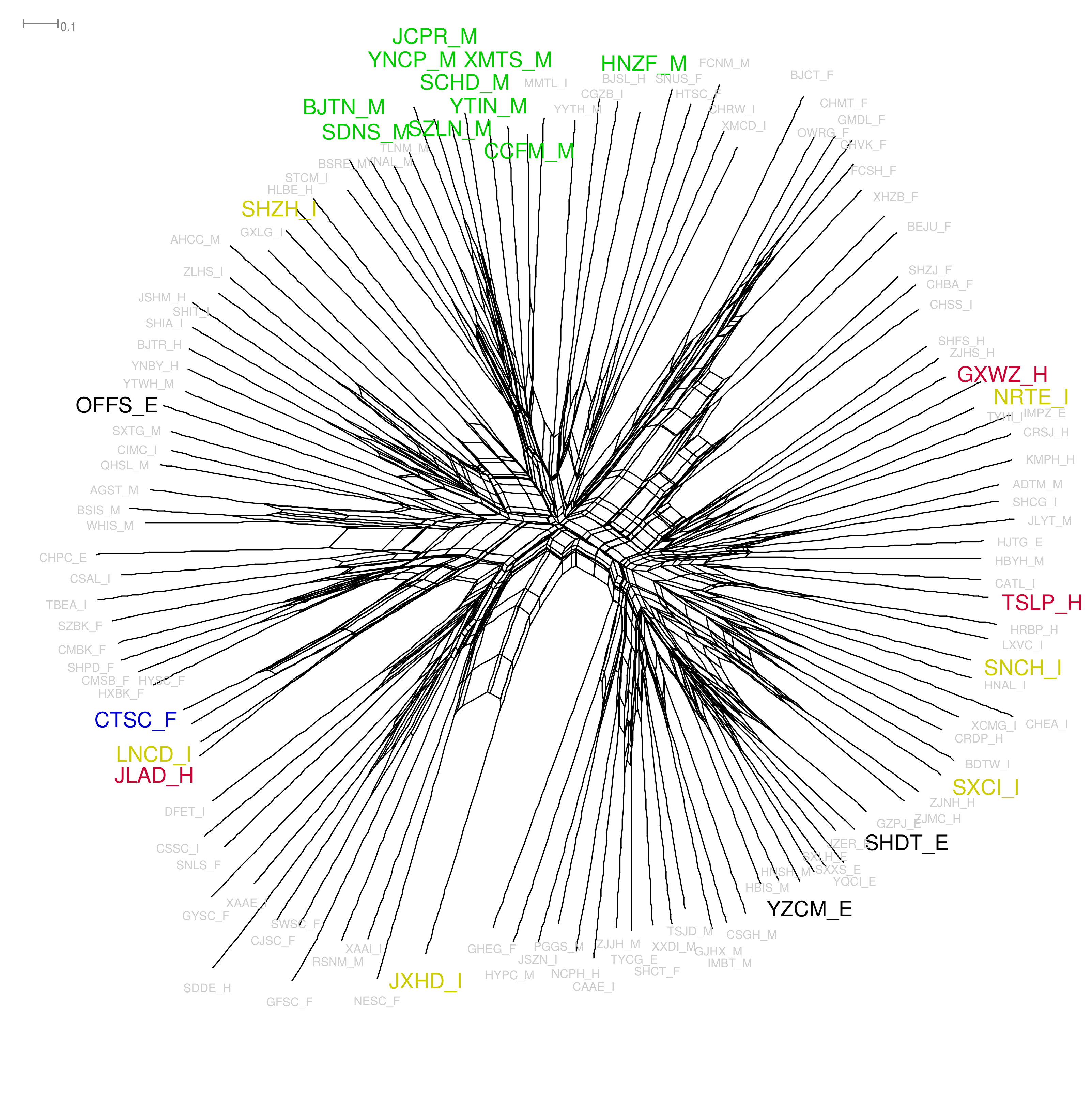}
  \caption{SplitsTree network
for study period two
with the stocks from cluster one, 
period one coloured. The colours are Energy - Black, Finance -- Blue,
Health Care -- Red, Industrials -- Khaki, Materials -- Green.  }
  \label{fig:ChinaP2Cl1P1}
\end{figure}
 
\begin{figure}[ht]
  \centering
  \includegraphics[width=12cm]{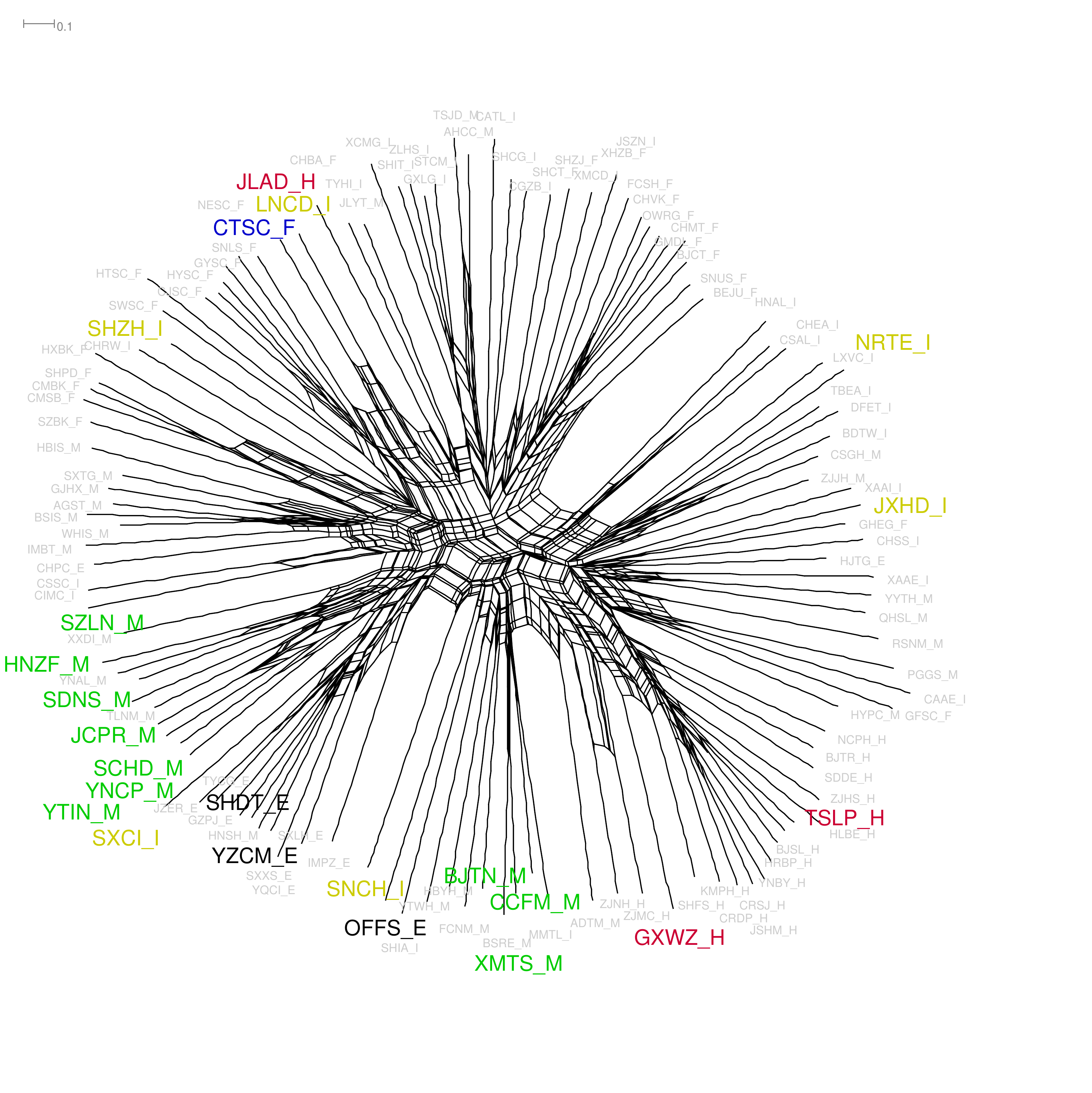}
  \caption{SplitsTree network
for study period three
with the stocks in cluster one,
period one coloured. The colours are Energy - Black, Finance -- Blue,
Health Care -- Red, Industrials -- Khaki, Materials -- Green.  }
  \label{fig:ChinaP3Cl1P1}
\end{figure}

\begin{figure}[ht]
  \centering
  \includegraphics[width=12cm]{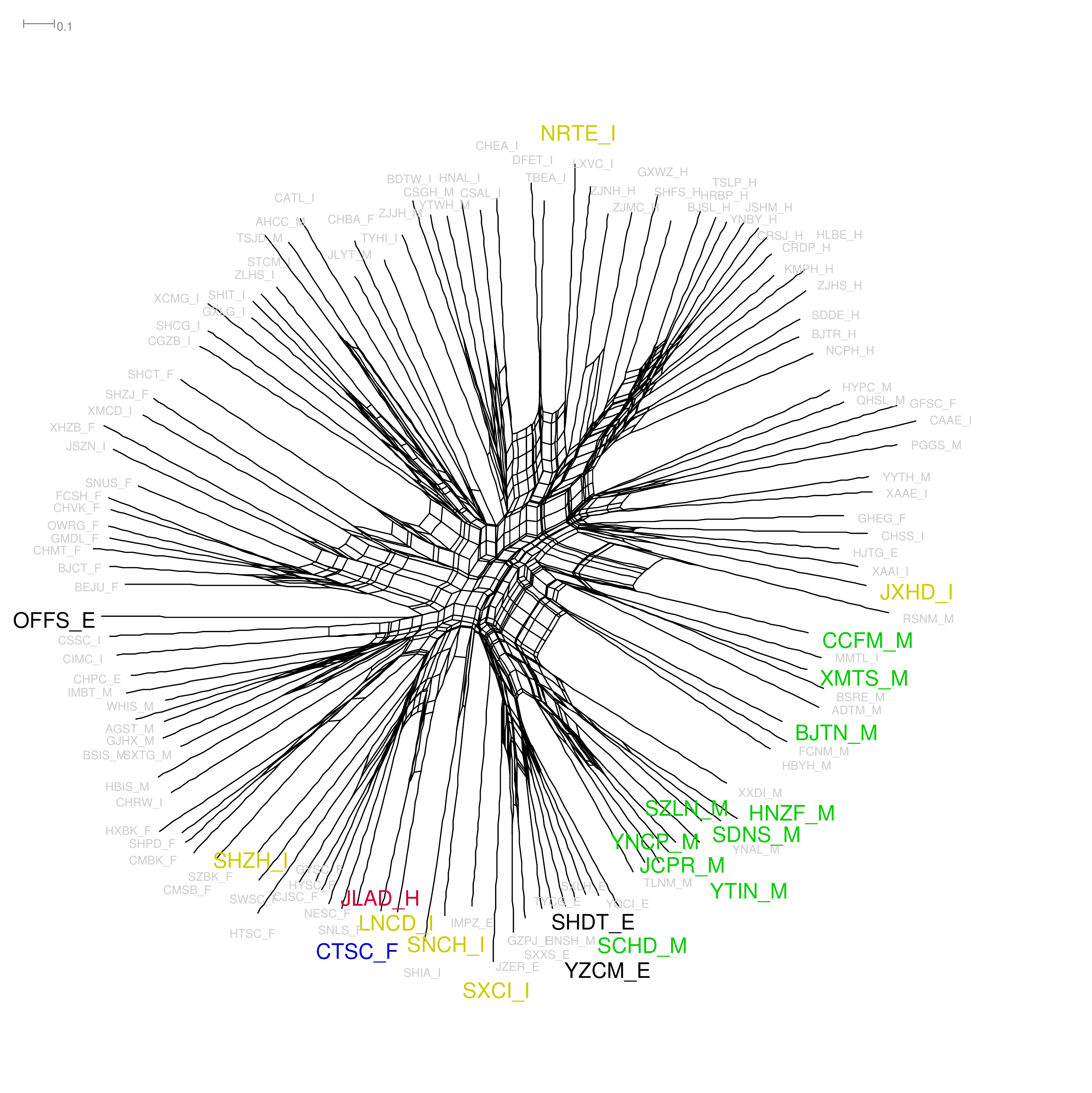}
  \caption{SplitsTree network
for study period four
with the stocks in cluster one,
period one coloured. The colours are Energy - Black, Finance -- Blue,
Health Care -- Red, Industrials -- Khaki, Materials -- Green.  }
  \label{fig:ChinaP4Cl1P1}
\end{figure}
 
\begin{figure}[ht]
  \centering
  \includegraphics[width=12cm]{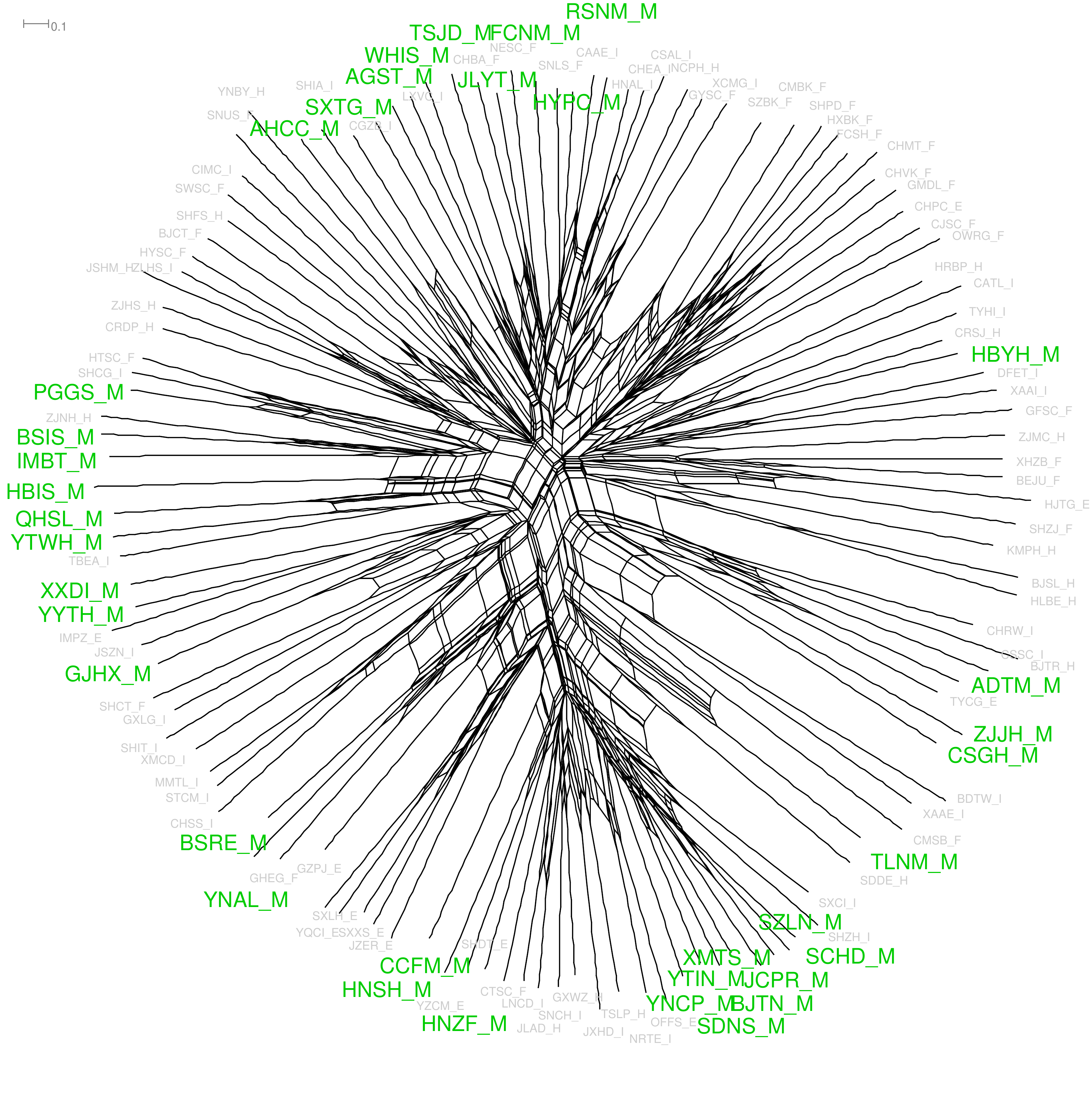}
  \caption{SplitsTree network
for study period one
with the stocks in the materials sector coloured green.}
  \label{fig:ChinaP1Materials}
\end{figure}
 
\begin{figure}[ht]
  \centering
  \includegraphics[width=12cm]{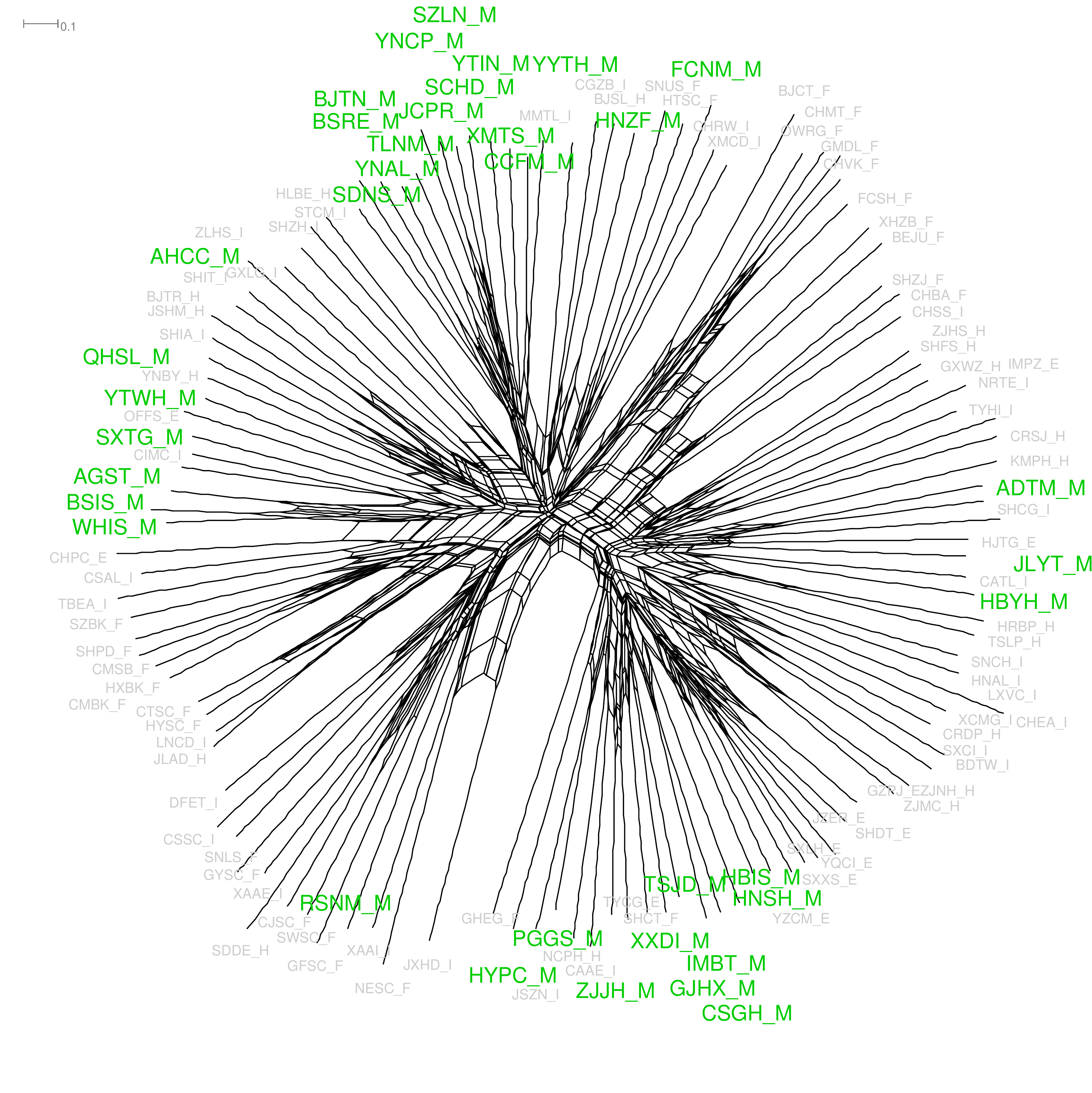}
  \caption{SplitsTree network
for study period two
with the stocks in the materials sector coloured green.}
  \label{fig:ChinaP2Materials}
\end{figure}
 
\begin{figure}[ht]
  \centering
  \includegraphics[width=12cm]{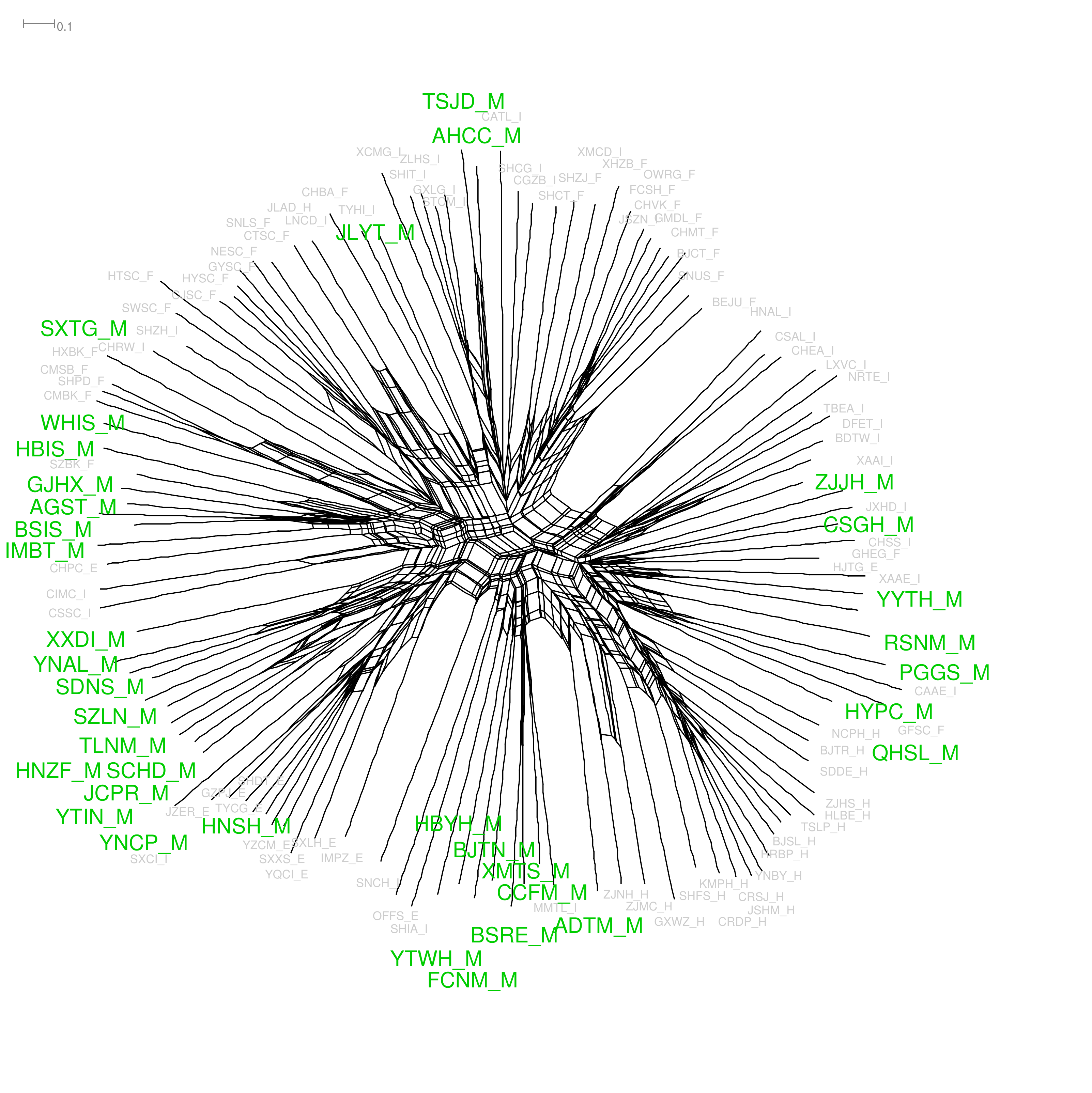}
  \caption{SplitsTree network
for study period three
with the stocks in the materials sector coloured green.}
  \label{fig:ChinaP3Materials}
\end{figure}
 
\begin{figure}[ht]
  \centering
  \includegraphics[width=12cm]{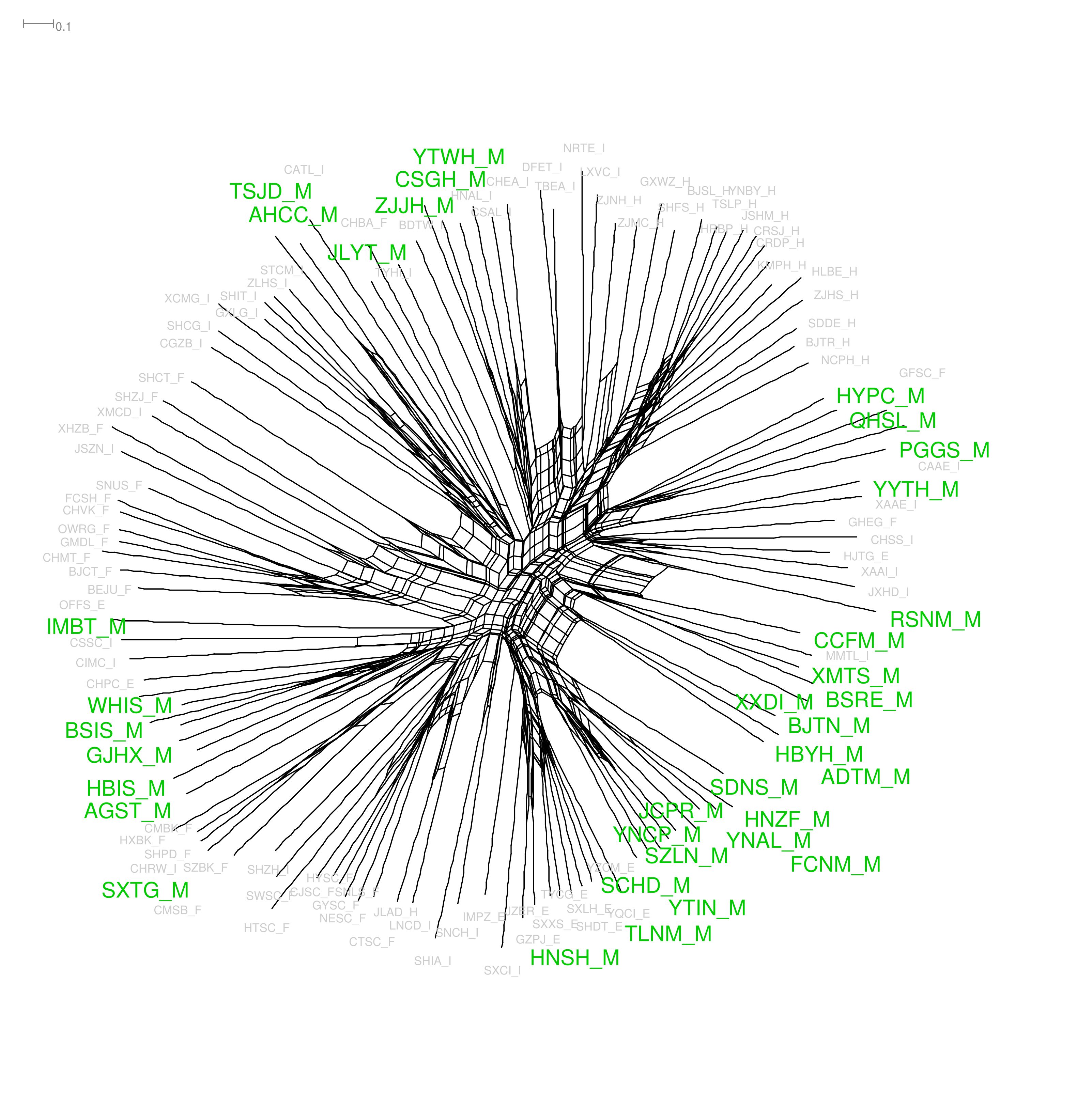}
  \caption{SplitsTree network
for study period four 
with the stocks in the materials sector coloured green.}
  \label{fig:ChinaP4Materials}
\end{figure}
 
%\begin{figure}[ht]
%  \centering
%  \includegraphics[width=12cm]{ChinaPeriod1-AllEnergy.pdf}
%  \caption{SplitsTree network
%for study period one 
%with the stocks in the energy sector coloured black.}
%  \label{fig:ChinaP1Energy}
%\end{figure}
 
%\begin{figure}[ht]
%  \centering
%  \includegraphics[width=12cm]{ChinaPeriod2-AllEnergy.pdf}
%  \caption{SplitsTree network
%for study period two 
%with the stocks in the energy sector coloured black.}
%  \label{fig:ChinaP2Energy}
%\end{figure}
 
%\begin{figure}[ht]
%  \centering
%  \includegraphics[width=12cm]{ChinaPeriod3-AllEnergy.pdf}
%  \caption{SplitsTree network
%for study period three
%with the stocks in the energy sector coloured black.}
%  \label{fig:ChinaP3Energy}
%\end{figure}
 
%\begin{figure}[ht]
%  \centering
%  \includegraphics[width=12cm]{ChinaPeriod4-AllEnergy.pdf}
%  \caption{SplitsTree network
%for study period four
%with the stocks in the energy sector coloured black.}
%  \label{fig:ChinaP4Energy}
%\end{figure}
 
In Figures (\ref{fig:ChinaP1Cl1CC}) through (\ref{fig:ChinaP4Materials})
we show the movement of industry groups both within a cluster and
 between study periods. We compare this with the movements
of the materials industry group in the splits graph.

In Figure 
(\ref{fig:ChinaP1Cl1CC}) we have selected Cluster 1 in study period
1 and assigned a colour to each industry group within the cluster.
While all five industry groups are represented in the cluster
it is clear that the materials group of stocks represent the
largest such group within this correlation cluster. Figures
(\ref{fig:ChinaP2Cl1P1}) through (\ref{fig:ChinaP4Cl1P1})
 shows locations
in the splits graph of the stocks from Cluster 1 of Period 1
in Periods 2 through 4. As can be seen the stocks in
this initial cluster do not remain clustered
together in subsequent periods. 

However, the materials group has remained together as a 
block not only in study period two but also in study periods three
and four. During period two (Figure \ref{fig:ChinaP2Cl1P1})
 the materials group
from Cluster 1 is now in what we identified as Cluster 3.
In study period three (Figure \ref{fig:ChinaP3Cl1P1}) they have
split into two groups and are in what we identified as
Clusters 1 and 6, which are adjacent clusters in that
study period. Finally in study period four they are
in what we identified as Clusters 1 and 2, again, these are
adjacent clusters in that study period.

In diversification one seeks groups of stocks which will tend
to move together in the future but relatively independently of
other so-identified groups of stocks. Then
an investors spreads their investments across
these groups. This is the basis for
previous studies which have grouped stocks by industry assuming
that stocks in the same industry will tend to have price
movements more similar than stocks in different industries,
see Section (\ref{sec:clusterindustry}) below.
Thus the evidence presented
here is that the stocks within Cluster one Period one from the
materials group form a financially useful grouping when 
forming a diversified portfolio for out-of-sample testing.

Because of this we would not expect
portfolios selected from stocks within correlation clusters alone
to be
significantly less risky than those chosen from industry groups.
%
% We can take the next bit out if we decide not to do it
% for this paper.
%
However, considering both a stock's industry group and its
correlation cluster has potential to result in greater risk
reduction than either method on its own.

\subsection{Clustering by Industry Group}\label{sec:clusterindustry}

In previous studies a number of authors have included in their
studies of forming
diversified stock portfolios at least one method in
which they dividied the stocks into
industry groups and then selected portfolios by spreading
the investments across the groups, see \cite{Domian2007}
for example. Neighbor-Nets splits graphs
give us a direct method of assessing the likely success of such
a strategy. To illustrate this we have selected the energy
and materials
groups because they had the smallest and
largest number of stocks, 12 and 36 respectively.
Figures (\ref{fig:ChinaP1Materials}) through
(\ref{fig:ChinaP4Materials}) show the locations of the 
materials stocks. Similar diagrams for the other
industry groups are available from the authors on request.

Clustering of the materials stocks is clearly visible in each
of the four study periods. This gives a direct visual confirmation
of previous studies which have reported that selecting stocks
by spreading them across industry groups gives a greater reduction 
in portfolio risk than randomly selecting stocks.

%% file: Example.tex
\section{Example}\label{sec:example}

This examples uses 126 stocks from the Shanghai exchange,
for which we
calculated the weekly returns from price and dividend data and
we divided the data into four periods based on market behaviour
as discussed in Section (\ref{sec:data}) above. Some basic statistics
on the correlations are presented in Table (\ref{tab:basicstats}).
As can be seen the highest average correlation occurred in period 3,
a time of a sharp market decline or crash.

For all the periods, as the portfolio size was increased the standard 
deviation of the returns decreased across all
four portfolio selection methods. 
Early empirical studies of portfolio diversification focused
on the number of stocks in a portfolio, see \cite{Evans1968}.
A larger portfolio was reported to be less risky with the lower risk 
being a result of the lower level of variation in the returns. However,
the benefit of reduced risk rapidly diminished with increasing
portfolio size.

An ANOVA test was used to compare the means, because
the variances were within a small range the ANOVA test remains valid even though the Levene test detects statistically significant differences. 
The Levene test was applied using the \verb+lawstat+ package in
\verb+R+ \citep{lawstat}.

\begin{table}[ht]
\begin{center}
\begin{tabular}{cccccc}
Period & Mean &Std. Dev.& Min & Max & Negative \\
\hline
1 & 0.266 & 0.170 & -0.642 & 0.864 & 438/7875 \\
2 & 0.328 & 0.196 & -0.413 & 0.855 & 480/7875 \\
3 & 0.441 & 0.191 & -0.168 & 0.908 & 132/7875 \\
4 & 0.437 & 0.192 & -0.158 & 0.906 & 143/7875 
\end{tabular}
\end{center}
\caption{Basic statistics on the correlations. There are
$n(n-1)/2=(126\times 125)/2=7875$ correlations between the
126 stocks. The final column gives the count of the number
of correlations which were estimated to be negative. The highest
proportion of negative correlations occurred in period
2 when approximately 6\% of estimated correlations were
negative.}\label{tab:basicstats}
\end{table}

\begin{table}[ht]
\begin{tabular}{cccccc}
Number of  &        &          &             & Industry and & ANOVA\\ 
Stocks & Random & Industry & Correlation & Correlation & (Levene) Test \\ 
in Portfolios & Selection & Grouping & Clusters & Clusters  & p-value\\ \hline
2 &   464 & 449   & 467   & 457    &  0.0783 \\
  & (234) & (227) & (220) &  (2.8) & (0.281) \\
4 &   468 & 459   & 463   & 4.71   & 0.248  \\ 
  & (169) & (161) & (154) & (158)  & (0.041) \\
8 &   466 & 459   & 454   & 4.64   & 0.484\\
  & (119) & (115) & (102) & (105)  & $(<$0.001)   \\
16&   466 & 462   & 463   & 466    & 0.023\\
  &   (78)& (78)  & (68)  & (50)   & $(<$0.001) 
\end{tabular}
\caption{Returns in percent
under the four different portfolio selection methods for period two using period one data for the estimation of the correlations. 
Underneath each set of returns, in brackets, is the standard deviation
of the returns. The final column reports the p-value of the ANOVA analysis which 
tests for differences in the means or the Levene test which
tests whether the standard deviations of all four methods are equal
as appropriate for each line. }\label{tab:period2returns}
\end{table}

Period two was a period of general market increase and the returns were good 
during this period. Table (\ref{tab:period2returns}) presents the mean and
standard deviations of returns together with some statistical
testing of the results.
The returns were statistically significantly different for portfolios of size 16 and weakly significant for portfolios of size 2.  For the smallest portfolios  the correlation cluster method performed best and for portfolios of size 4 and 16 the industry and correlation clusters method performed best. 

For all the portfolios the variation in the returns decreased as the portfolio size increased. The Levene test showed that there was statistically significant differences in the 
standard deviations for portfolios of size 4, 8 and 16. For portfolios of 
size  4 and 8 the correlation cluster method produced the lowest variation in 
the returns. For portfolios of size 16 it was the industry and correlation 
cluster method that produces the lowest variation, by a substantial margin.

\begin{table}[ht]
\begin{tabular}{cccccc }
Number of  &        &          &             & Industry and  & ANOVA\\ 
Stocks & Random & Industry & Correlation &Correlation & (Levene) Test \\ 
in Portfolios & Selection & Grouping & Clusters & Clusters  &p-value \\ \hline
2  & -57    & -54  & -52  & -53  & 0.007  \\
   & (25)   & (27) & (29) & (27) & (0.265) \\
4  & -58    & -56  & -53  & -54  & 0.001   \\   
   & (0.16) & (17) & (19) & (18) & (0.001) \\
8  & -57    & -55  & -53  & -54  & $<$0.001 \\   
   & (0.11) & (12) & (14) & (13) & (0.004) \\
16 & -57    & -54  & -55  & -54  & $<$0.001 \\
   & (8)    & (8)  & (8)  & (7)  & $(<$0.001) \\
\end{tabular}
\caption{Returns in percent
under the four different portfolio selection methods for period three  using period two data for the estimation of the correlations.
Underneath each set of returns, in brackets, is the standard deviation
of the returns. The final column reports the p-value of the ANOVA analysis which 
tests for differences in the means or the Levene test which
tests whether the standard deviations of all four methods are equal as 
appropriate.}\label{tab:period3returns}
\end{table}

Table (\ref{tab:period3returns}) presents the mean and standard deviations of 
returns together with some statistical testing of the results for period
three. This
was a period of general market decline. In
these circumstances a widely used risk/return measure such as the
Sharpe ratio is negative. In such circumstances a  private
investor would regard a portfolio which minimised the losses as 
be the most desirable. While we should not over interpret the
results, the correlation clusters have slightly better returns for 
portfolios of sizes 2, 4 and 8. The industry and correlation clusters and industry based groupings have slightly better returns for portfolios of size 16. 

As with period two out of sample testing, the variation decreased as 
the portfolio size was increased, regardless of the method used to 
select the portfolio. The Levene test showed that there was statistically 
significant differences in the variances in the standard deviations for 
portfolios of size 4, 8, and 16.  Typically the correlation cluster 
method showed the largest standard deviations and random selection method the lowest standard deviations. For portfolios of size 16 the industry and correlation clusters method reported the smallest variation.

\begin{table}[ht]
\begin{tabular}{cccccc }
Number of & & & & Industry and & ANOVA\\ 
Stocks & Random & Industry & Correlation & Correlation & (Levene) test \\ 
in Portfolios & Selection & Grouping & Clusters & Clusters & p-value \\ \hline
2  &2.2    & 211    & 241   & 237    &   $<$0.001\\
   & (154) & (164)  & (173) & (166)  &(0.227)  \\
4  & 229   & 200    & 235   & 233    & $<$0.001 \\ 
   & (118) &  (105) & (113) & (118)  & $(<$0.001)  \\
8  & 218   & 210    & 233   & 234   & $<$0.001\\
   & (75)  & (74)   & (82)  & (84)  &(0.003)  \\
16 & 219   & 207    & 232   & 234   & $<$0.001 \\
   & (53)  & (50)   & (53)  & (41)  &$(<$0.001)  
\end{tabular}
\caption{Returns in percent 
under the four different portfolio selection methods for period four  using period three data for the estimation of the correlations.
Underneath each set of returns, in brackets, is the standard deviation
of the returns. The final column reports the p-value of the ANOVA analysis which 
tests for differences in the means or the Levene test which
tests whether the standard deviations of all four methods are equal
as appropriate.} \label{tab:period4returns} 
\end{table}

Table (\ref{tab:period4returns}) presents the mean and standard deviations of returns together with some statistical testing of the results for period four.
This period showed modest returns. While, again, we should not
over-interpret the results, the returns were lower for random and 
industry grouping selection methods for all four portfolios sizes tested. 
The highest returns were for the correlation clusters portfolio selection 
method for the two smaller portfolios, and for portfolios of size 8 and 16 the industry and correlation clusters method reports slightly higher returns. 

As with period two and three out of sample testing, the variation decreased 
as the portfolio size was increased, regardless of the method used to 
select the portfolio. The Levene test showed that there was statistically 
significant differences in the variances for the portfolios of sizes 4, 
8 and 16. The industry based selection method offered the greatest reduction 
in the variation in the returns for portfolios of size 4 and 8. For the largest portfolio size (portfolios of size 16) the industry and correlation clusters had the lowest standard deviations (the same outcome as periods two and three). 

Therefore this suggests that the correlation clusters (or industry and correlation clusters) are particularly effective in times of general market increase, with the benefit being either a reduction in the variation  or an increase in the return. 

This study shows that combining industry and correlation clusters is particularly effective at lowering the variation for the larger portfolios, with all three periods showing a much lower variation for portfolios of size 16, as well as reasonable returns. This is in line with general advice to investors to hold larger portfolios and to ensure the holdings are diversified. 

%The portfolios selected according to correlations clusters and industry groupings were the best performers for the smaller portfolios (those of size two or four). For the two larger portfolio sizes (eight and 16) the industry grouping portfolio selection method outperformed the portfolios selected according to correlation clusters and industry groupings.

%% file: Discussion.tex
\section{Discussion}\label{sec:discussion}

An earlier paper \citep{Rea2014} introduced Neighbor-Net networks as a method for visualising correlations in stock markets. The method has the advantage of being able to represent a lot of the key features of the correlation matrix in a planar graphic. The paper noted that such a diagram could assist with creating diversified portfolios. This paper has highlighted the effectiveness of using correlation clusters to investigate diversified portfolios. 

In this paper four risk budgeting
methods of portfolio selection were compared; randomly selected portfolios, industry clusters, correlation clusters and industry and correlation clusters. Traditionally selecting stocks by industry was considered an appropriate method to diversify a portfolio. While this may be the case in some markets and under some market conditions, this investigation demonstrated that industry based clusters was generally outperformed by portfolios selected at random, however the portfolios selected using industry grouping may have lower variance in times of market increase compared with random selection. 

Of the four, the most restrictive method of selecting portfolios was the 
industry and correlation cluster selection method. With the random selection method all possible combinations of $n$ stocks from the 126 stocks are allowable but for the industry and correlation cluster selection method, there are many portfolios that are not admissible because they do not meet the rules of this portfolio selection method. The industry grouping and correlation cluster methods are also restrictive but less so than the industry and correlation clusters method. 

The main concern was whether the rules of portfolio selection presented here 
offer significant benefits. If a difference in mean was detected, the correlation clusters or industry and correlation clusters method may outperform the other methods on mean return. This effect was most pronounced in the period four out of sample testing where the returns for the correlations clusters and industry and correlation clusters method always exceeded random portfolio selection. Therefore the knowledge of the circular ordering can be used to enhance portfolio returns. 

The variation in the returns for portfolios of size 16 was always lowest if the method of portfolio selection was Industry and Correlation Cluster selection. For the other portfolio sizes the variation with a method decreases as the portfolio size increases, but no one method consistently outperforms the others. This suggests portfolio size has a greater impact on the variation of the returns than the method used to select of the portfolio.

\cite{Rea2014} discussed how stocks from the opposite side of the Neighbor-Net network did not necessarily create a portfolio with high returns
because some stocks maybe giving negative returns while one on the
opposite side of the network may be giving positive returns. Dividing the data into four periods in the manner we did,
represents a particularly severe
test of diversification, particularly since no account was taken of
either historical or expected returns of the stocks.
It is 
our expectation that investor knowledge and analysis alongside correlation cluster based portfolio selection has the potential to improve the return of the portfolio, as well as reduce the variance (or equivalently, the standard deviation). But this awaits further research.

We note that the correlation clusters were determined by eye in this analysis. 
This is a valid method of determining clusters, exploiting both the
structure of the network and the circular ordering of stocks the
neighbor-Net algorithm produces.
Future work could focus on methods to automate the selection  of the 
correlation clusters to see if this further enhances the portfolio performance.

%% file: 126Companynames.tex
\section{Stock Codes and Industry Segments}\label{sec:stockcodes}

\begin{longtable}{lll}
\caption[stock codes]{Stock market codes and company names and 
Industrial sector of stocks
in the study.} \\ 
\hline
Company Name &Company Code&Industry Group \\
\hline
\endhead
\hline
\endfoot
China Ptl. \& Chm.&CHPC-E&Energy \\
Guizhou Panjiang Coal&GZPJ-E&Energy \\
Inner Mongolia Pingzhuang En. Rso.&IMPZ-E&Energy \\
Jizhong Energy Res.&JZER-E&Energy \\
Liaoning Hjtg. Chems. & HJTG-E & Energy \\
Offs. Oil Engr.&OFFS-E&Energy \\
Shai Datun Energy Res. &SHDT-E&Energy \\
Shanxi Lanhua Sci-Tech Venture&SXLH-E&Energy \\
Shanxi Xishan&SXXS-E&Energy \\
Taiyuan Coal Gasification&TYCG-E&Energy \\
Yangquan Coal &YQCI-E&Energy \\
Yanzhou Coal Mining&YZCM-E&Energy \\
\hline
Beijing Capital Dev.&BJCT-F&Finance \\
Bej. Urban Con. Inv. Dev. &BEJU-F&Finance \\
Changjiang Securities &CJSC-F&Finance \\
China Baoan Gp. & CHBA-F&Finance \\
China Merchants Bank &CMBK-F&Finance \\
China Merchants Pr. Dev.&CHMT-F&Finance \\
China Minsheng Banking &CMSB-F&Finance \\
China Vanke &CHVK-F&Finance \\
Citic Securities &CTSC-F&Finance \\
Financial Str. Sldg.&FCSH-F&Finance \\
Gemdale &GMDL-F&Finance \\
GF Securities &GFSC-F&Finance \\
Guanghui Energy &GHEG-F&Finance \\
Guoyuan Securities &GYSC-F&Finance \\
Haitong Securities &HTSC-F&Finance \\
Hong Yuan Secs. &HYSC-F&Finance \\
Huaxia Bank &HXBK-F&Finance \\
Northeast Securities &NESC-F&Finance \\
Oceanwide Rlst. Group &OWRG-F&Finance \\
Shai. Chengtou Hldg.&SHCT-F&Finance \\
Shai. Pudong Dev. Bk. &SHPD-F&Finance \\
Shai. Zhangjiang &SHZJ-F&Finance \\
Shenzhen Dev. Bank&SZBK-F&Finance \\
Sinolink Securities &SNLS-F&Finance \\
Southwest Securities &SWSC-F&Finance \\
Suning Universal &SNUS-F&Finance \\
Xinhu Zhongbao&XHZB-F&Finance \\
\hline
Beijing Sl Pharmaceutical &BJSL-H&Health Care  \\
Beijing Tongrentang &BJTR-H&Health Care  \\
China Res. Dble. Crane Pharm.&CRDP-H&Health Care  \\
China Res. Sanjiu Med.\& pharm.&CRSJ-H&Health Care  \\
Guangxi Wuzhou Zhongheng &GXWZ-H&Health Care  \\
Harbin Pharms. Gp. &HRBP-H&Health Care  \\
Hualan Biological Engr. &HLBE-H&Health Care  \\
Jiangsu Hengrui Medicine &JSHM-H&Health Care  \\
Jilin Aodong Pharm. Gp. &JLAD-H&Health Care  \\
Kangmei Pharm.&KMPH-H&Health Care  \\
North China Pharm. &NCPH-H&Health Care  \\
Shai. Fosun Pharm. Group &SHFS-H&Health Care  \\
Shan Dong Dong E-Jiao &SDDE-H&Health Care  \\
Tasly Pharmaceutical &TSLP-H&Health Care  \\
Yunnan Baiyao Gp.&YNBY-H&Health Care  \\
Zhejiang Hisun Pharm. &ZJHS-H&Health Care  \\
Zhejiang Medicine &ZJMC-H  &Health Care  \\
Zhejiang Nhu &ZJNH-H&Health Care  \\
\hline
Baoding Tianwei Baobian Elec.&BDTW-I&Industrial \\
China Avic Avionics Equ. &CAAE-I&Industrial \\
China Cssc Hdg.&CSSC-I&Industrial \\
China Eastern Airl. &CHEA-I&Industrial \\
China Gezhouba Group &CGZB-I&Industrial \\
China Intl.Mar.Ctrs.&CIMC-I &Industrial \\
China Railway Erju &CHRW-I&Industrial \\
China Railway Tielong Container Logistic&CATL-I& Industrial \\
China Southern Airlines &CSAL-I&Industrial \\
China Spacesat &CHAA-I&Industrial \\
Dongfang Electric &DFET-I&Industrial \\
Guangxi Liugong Mch&GXLG-I&Industrial \\
Hainan Airlines &HNAL-I&Industrial \\
Jiangsu Zhongnan Con.&JSZN-I&Industrial \\
Jiangxi Hongdu Aviation &JXHD-I&Industrial \\
Liaoning Chengda &LNCD-I&Industrial \\
Luxin Venture Cap. Gp.&LXVC-I&Industrial \\
Minmetals Dev.&MMTL-I&Industrial \\
Nari Tech. Dev. &NRTE-I&Industrial \\
Sany Heavy Industry &SHIT-I&Industrial \\
Shai. Shenhua Heavy Ind. &SHZH-I&Industrial \\
Shanghai Con. Group&SHCG-I&Industrial \\
Shanghai Intl. Arpt.&SHIA-I&Industrial \\
Shantui Con. Mch.&STCM-I&Industrial \\
Shanxi Coal Intl. &SXCI-I&Industrial \\
Sinochem Intl. &SNCH-I&Industrial \\
Taiyuan Hvy. Ind.&TYHI-I&Industrial \\
Tbea &TBEA-I&Industrial \\
Xcmg Con. Machinery&XCMG-I&Industrial \\
Xi'an Aero-Engine &XAAE-I&Industrial \\
Xi'an Air.Intl.&XAAI-I&Industrial \\
Xiamen C \& D&XCMD-I&Industrial \\
Zoomlion Hdy. Sctc.&ZLHS-I&Industrial \\
\hline
Advd. Tech.\& Mats. &ADTM-M&Materials \\
Angang Steel &AGST-M&Materials \\
Anhui Conch Cmt. &AHCC-M&Materials \\
Baoji Titanium Ind.&BJTN-M&Materials \\
Baoshan Iron \& Stl.&BSIS-M&Materials \\
China Nonferrous Mtl.&CCFM-M&Materials \\
Csg Holding  &CSGH-M&Materials \\
Fangda Cbn. New Mra. &FCNM=M&Materials \\
Gan Jiu Stl. Gp. Hongxing &GJHX-M&Materials \\
Ginghai Salt Lake Ind. &QHSL-M&Materials \\
Industrial Sichuan Hongda&SCHD-M&Materials \\
Inmong. Baotou Stl. Rare Earth &BSRE-M&Materials \\
Hebei Iron \& Steel &HBIS-M&Materials \\
Henan Shenhuo Caa. \& Pwr. &HNSH-M&Materials \\
Henan Zhongfu Indl.&HNZF-M&Materials \\
Hengyi Petrochemical &HYPC-M&Materials \\
Hubei Yihua Chm. Ind.&HBYH-M&Materials \\
Inner Mongolia Baotou Steel Union &IMBT-M&Materials \\
Jiangxi Cpr. &JCPR-M&Materials \\
Jilin Yatai Group &JLYT-M&Materials \\
Pangang Gp. Stl. Vmtm.&PGGS-M&Materials \\
Rising Nonfr. Mtls&RSNM-M&Materials \\
Shandong Nanshan Almn.&SDNS-M&Materials \\
Shanxi Taigang Stl. &SXTG-M&Materials \\
Shn. Zhongjin Lingnan Nonfemet &SZLN-M&Materials \\
Tangshan Jidong Cmt.&TSJD-M&Materials \\
Tongling Nonfr. Mtls. Gp. &TLNM-M&Materials \\
Xiamen Tungsten &XMTS-M&Materials \\
Xinxing Ductile Iron &XXDI-M&Materials \\
Yantai Wanhua Polyuretha &YTWH-M&Materials \\
Yunnan Alum. &YNAL-M&Materials \\
Yunnan Copper &YNCP-M&Materials \\
Yunnan Tin &YTIN-M&Materials \\
Yunnan Yuntianhua &YYTH-M&Materials \\
Wuhan Iron and Steel &WHIS-M&Materials \\
Zhejiang Juhua &ZJJH-M&Materials 
\label{tab:stockcodes}
\end{longtable}

%% file: Period1Clusters.tex
\section{Stocks in Each Cluster}\label{sec:stockclusters}

\subsection{Period 1}\label{sub:period1}
\begin{description}
\item[Cluster1:] YZCM\_E, SXCI\_I, OFFS\_E, SCHD\_M, JCPR\_M, SHZH\_I, YNCP\_M, JXHD\_I, CCFM\_M, BJTN\_M, GXWZ\_H, SNCH\_I, XMTS\_M, YTIN\_M, HNZF\_M, JLAD\_H, TSLP\_H, LNCD\_I, NRTE\_I, CTSC\_F, SZLN\_M, SHDT\_E, SDNS\_M
\item[Cluster2:]  XAAE\_I, CSSC\_I, TYCG\_E, CSGH\_M, ADTM\_M, SDDE\_H, CMSB\_F, BDTW\_I, CHRW\_I, BJTR\_H, TLNM\_M, ZJJH\_M
\item[Cluster3:] BEJU\_F, KMPH\_H, ZJMC\_H, HJTG\_E, SHZJ\_F, XAAI\_I, TYHI\_I, CATL\_I, BJSL\_H, CRSJ\_H, DFET\_I, HLBE\_H, XHZB\_F, HBYH\_M, HRBP\_H, GFSC\_F
\item[Cluster4:] CMBK\_F, CJSC\_F, OWRG\_F, HXBK\_F, CHMT\_F, FCSH\_F, SZBK\_F, GMDL\_F, CHPC\_E, SHPD\_F, CHVK\_F
\item[Cluster5:]  JLYT\_M, CSAL\_I, SNUS\_F, NESC\_F, AHCC\_M, ZLHS\_I, HNAL\_I, XCMG\_I, GYSC\_F, SXTG\_M, BJCT\_F, ZJHS\_H, JSHM\_H, CRDP\_H, CGZB\_I, FCNM\_M, SNLS\_F, TSJD\_M, YNBY\_H, WHIS\_M, SHFS\_H, CHEA\_I, CAAE\_I, HYPC\_M, CHBA\_F, SWSC\_F, HYSC\_F, CIMC\_I, AGST\_M, RSNM\_M, SHIA\_I, NCPH\_H, LXVC\_I
\item[Cluster6:]  BSIS\_M, HTSC\_F, HBIS\_M, SHCG\_I, PGGS\_M, IMBT\_M, TBEA\_I, QHSL\_M, ZJNH\_H, YTWH\_M
\item[Cluster7:]  YYTH\_M, XMCD\_I, CHSS\_I, JSZN\_I, XXDI\_M, GXLG\_I, SHIT\_I, SHCT\_F, STCM\_I, GJHX\_M, IMPZ\_E, MMTL\_I
\item[Cluster8:]  YNAL\_M, SXLH\_E, BSRE\_M, JZER\_E, SXXS\_E, YQCI\_E, GHEG\_F, HNSH\_M, GZPJ\_E
\end{description}

%% file: StockNNetDiversification.bbl
\begin{thebibliography}{}

\bibitem[\protect\citeauthoryear{Bai and Green}{Bai and Green}{2010}]{Bai2010}
Bai, Y. and C.~J. Green (2010).
\newblock {International Diversification Strategies: Revisited from the Risk
  Perspective}.
\newblock {\em The Journal of Banking and Finance\/}~{\em 34}, 236--245.

\bibitem[\protect\citeauthoryear{Bali, Cakici, Yan, and Zhang}{Bali
  et~al.}{2005}]{Bali2005}
Bali, T.~G., N.~Cakici, X.~Yan, and Z.~Zhang (2005).
\newblock {Does Idiosyncratic Risk Really Matter?}
\newblock {\em The Journal of Finance\/}~{\em 60\/}(2), 905--929.

\bibitem[\protect\citeauthoryear{Barber and Odean}{Barber and
  Odean}{2008}]{Barber2008}
Barber, B.~M. and T.~Odean (2008).
\newblock {All That Glitters: The Effect of Attention and News on the Buying
  Behaviour of Individual and Institutional Investors}.
\newblock {\em The Review of Financial Studies\/}~{\em 21\/}(2), 785--818.

\bibitem[\protect\citeauthoryear{Benzoni, Collin-Dufresne, and
  Goldstein}{Benzoni et~al.}{2007}]{Benzoni2007}
Benzoni, L., P.~Collin-Dufresne, and R.~S. Goldstein (2007).
\newblock {Portfolio Choice over the Life-Cycle when the Stock and Labor
  Markets are Cointegrated}.
\newblock {\em The Journal of Finance\/}~{\em 62\/}(5), 2123--2167.

\bibitem[\protect\citeauthoryear{Bergqvist, Forsman, andJ. Naslund, Lilga,
  Engdahl, Lindstrom, Gylfe, Ahlm, Evanders, and Bucht}{Bergqvist
  et~al.}{2015}]{Bergqvist2015}
Bergqvist, J., O.~Forsman, P.~L. andJ. Naslund, T.~Lilga, C.~Engdahl,
  A.~Lindstrom, A.~Gylfe, C.~Ahlm, M.~Evanders, and G.~Bucht (2015).
\newblock Detection and isolation of sindbis virus from mosquitoes captured
  during an outbreak in sweded.
\newblock {\em Vector-borne and Zoonotic Diseases\/}~{\em 15\/}(2), 133--140.

\bibitem[\protect\citeauthoryear{Bryant and Moulton}{Bryant and
  Moulton}{2004}]{Bryant2004}
Bryant, D. and V.~Moulton (2004).
\newblock Neighbor-net: An agglomerative method for the construction of
  phylogenetic networks.
\newblock {\em Molecular Biology and Evolution\/}~{\em 21\/}(2), 255--265.

\bibitem[\protect\citeauthoryear{Cont}{Cont}{2001}]{Cont2001}
Cont, R. (2001).
\newblock {Empirical properties of asset returns: stylized facts and
  statistical issues}.
\newblock {\em Quantitative Finance\/}~{\em 1:2}, 223--236.

\bibitem[\protect\citeauthoryear{Domian, Louton, and Racine}{Domian
  et~al.}{2007}]{Domian2007}
Domian, D.~L., D.~A. Louton, and M.~D. Racine (2007).
\newblock {Diversification in Portfolios of Individual Stocks: 100 Stocks Are
  Not Enough}.
\newblock {\em The Financial Review\/}~{\em 42}, 557--570.

\bibitem[\protect\citeauthoryear{Emadzade, Lebmann, Hoffmann, Tkach, Lone, and
  Horandi}{Emadzade et~al.}{2015}]{Emadzade2015}
Emadzade, K., M.~J. Lebmann, M.~H. Hoffmann, N.~Tkach, F.~A. Lone, and
  E.~Horandi (2015).
\newblock Phylogenetic relationships and evolution of high mountain buttercups
  (ranunculus) in north america and central asia.
\newblock {\em Perspectives in Plant Ecology, Evolution and Systematics\/}~{\em
  17}, 131--141.

\bibitem[\protect\citeauthoryear{Evans and Archer}{Evans and
  Archer}{1968}]{Evans1968}
Evans, J.~L. and S.~H. Archer (1968).
\newblock {Diversification and the Reduction of Dispersion: An Empirical
  Analysis}.
\newblock {\em The Journal of Finance\/}~{\em 23\/}(5), 761--767.

\bibitem[\protect\citeauthoryear{Fama and French}{Fama and
  French}{1992}]{Fama1992}
Fama, E.~F. and K.~R. French (1992).
\newblock {The Cross-Section of Expected Stock Returns}.
\newblock {\em The Journal of Finance\/}~{\em 47\/}(2), 427--465.

\bibitem[\protect\citeauthoryear{French and Fama}{French and
  Fama}{1993}]{Fama1993}
French, K.~R. and E.~F. Fama (1993).
\newblock {Common Risk Factors in the Returns on Stocks and Bonds}.
\newblock {\em Journal of Financal Economics\/}~{\em 33}, 3--56.

\bibitem[\protect\citeauthoryear{Gastwirth, Gel, Hui, Lyubchich, Miao, and
  Noguchi}{Gastwirth et~al.}{2013}]{lawstat}
Gastwirth, J.~L., Y.~R. Gel, W.~L.~W. Hui, V.~Lyubchich, W.~Miao, and
  K.~Noguchi (2013).
\newblock {\em lawstat: An R package for biostatistics, public policy, and
  law}.
\newblock R package version 2.4.1.

\bibitem[\protect\citeauthoryear{Goyal and Santa-Clara}{Goyal and
  Santa-Clara}{2003}]{Goyal2003}
Goyal, A. and P.~Santa-Clara (2003).
\newblock {Idiosyncratic Risk Matters!}
\newblock {\em The Journal of Finance\/}~{\em 58\/}(3), 975--1007.

\bibitem[\protect\citeauthoryear{Gray, Bryant, and Greenhill.}{Gray
  et~al.}{2010}]{Grey2010}
Gray, R., D.~Bryant, and S.~J. Greenhill. (2010).
\newblock On the shape and fabric of human history.
\newblock {\em Philosophical Transactions of the Royal Society B\/}~{\em 365},
  3923--3933.

\bibitem[\protect\citeauthoryear{Heggarty, Maguire, and McMahon}{Heggarty
  et~al.}{2010}]{Heggarty2010}
Heggarty, P., W.~Maguire, and A.~McMahon (2010).
\newblock Splits or waves? trees or webs? how convergence measures and network
  analysis can unravel language histories.
\newblock {\em Philosophical Transactions of the Royal Society B\/}~{\em 365},
  3829--3843.

\bibitem[\protect\citeauthoryear{Huson and Bryant}{Huson and
  Bryant}{2006}]{Huson2006}
Huson, D.~H. and D.~Bryant (2006).
\newblock Application of phylogenetic networks in evolutionary studies.
\newblock {\em Molecular Biology and Evolution\/}~{\em 23\/}(2), 254--267.

\bibitem[\protect\citeauthoryear{Jordan and O'Neil}{Jordan and
  O'Neil}{2010}]{Jordan2010}
Jordan, P. and S.~O'Neil (2010).
\newblock Untangling cultural inheritance: language diversity and long-house
  architecture on the pacific northwest coast.
\newblock {\em Philosophical Transactions of the Royal Society B\/}~{\em 365},
  3875--3888.

\bibitem[\protect\citeauthoryear{Jorion}{Jorion}{1985}]{Jorion1985}
Jorion, P. (1985).
\newblock {International Portfolio Diversification with Estimation Risk}.
\newblock {\em The Journal of Business\/}~{\em 58\/}(3), 259--278.

\bibitem[\protect\citeauthoryear{Knooihuizen and Dediu}{Knooihuizen and
  Dediu}{2012}]{Knooihuizen2012}
Knooihuizen, R. and D.~Dediu (2012).
\newblock Historical demography and historical sociolinguistics: The role of
  migrant integration in the development of dunkirk french in the 17th century.
\newblock {\em Language Dynamics and Change\/}~{\em 2\/}(1), 1--33.

\bibitem[\protect\citeauthoryear{Lee}{Lee}{2011}]{Lee2011}
Lee, W. (2011).
\newblock {Risk-Based Asset Allocation: A New Answer to an Old Question?}
\newblock {\em Journal of Portfolio Management\/}~{\em 37\/}(4), 11--28.

\bibitem[\protect\citeauthoryear{Lowenfeld}{Lowenfeld}{1909}]{lowenfeld1909}
Lowenfeld, H. (1909).
\newblock {\em {Investment, an Exact Science}}.
\newblock Financial Review of Reviews.

\bibitem[\protect\citeauthoryear{Mantegna}{Mantegna}{1999}]{Mantegna1999}
Mantegna, R.~N. (1999).
\newblock {Hierarchical structure in financial markets}.
\newblock {\em The European Physical Journal B\/}~{\em 11}, 193--197.

\bibitem[\protect\citeauthoryear{Markowitz}{Markowitz}{1991}]{markowitz1959}
Markowitz, H.~M. (1991).
\newblock {\em {Portfolio Selection: Efficient Diversification of Investments
  2nd Edition}}.
\newblock Wiley.

\bibitem[\protect\citeauthoryear{Markowtiz}{Markowtiz}{1952}]{markowitz1952}
Markowtiz, H. (1952).
\newblock {Portfolio Selection}.
\newblock {\em The Journal of Finance\/}~{\em 7\/}(1), 77--91.

\bibitem[\protect\citeauthoryear{Prentiss, Skelton, Eldredge, and
  Quinn}{Prentiss et~al.}{2011}]{Prentiss2011}
Prentiss, A.~M., R.~R. Skelton, N.~Eldredge, and C.~Quinn (2011).
\newblock Get rad! the evolution of the skateboard deck.
\newblock {\em Evo Edu Outreach\/}~{\em 4}, 379--389.

\bibitem[\protect\citeauthoryear{Rea and Rea}{Rea and Rea}{2014}]{Rea2014}
Rea, A. and W.~Rea (2014).
\newblock {Visualization of a stock market correlation matrix}.
\newblock {\em Physica A\/}~{\em 400}, 109--123.

\bibitem[\protect\citeauthoryear{Ross, Greenhill, and Atkinson}{Ross
  et~al.}{2013}]{Ross2013}
Ross, R.~M., S.~J. Greenhill, and Q.~D. Atkinson (2013).
\newblock Population structure and cultural geography of a folktale in europe.
\newblock {\em Proceedings of the Royal Society B\/}~(280), 1756.

\bibitem[\protect\citeauthoryear{Schwarz, Ng, Cooke, Newman, Temple, Piskorz,
  Gale, Sayal, Murtaza, Baldwin, Rosenfeld, Earl, Sala, Jimenez-Linan,
  Parkinson, Markowetz, and Brenton}{Schwarz et~al.}{2015}]{Schwarz2015}
Schwarz, R.~F., C.~K.~Y. Ng, S.~L. Cooke, S.~Newman, J.~Temple, A.~M. Piskorz,
  D.~Gale, K.~Sayal, M.~Murtaza, P.~J. Baldwin, N.~Rosenfeld, H.~M. Earl,
  E.~Sala, M.~Jimenez-Linan, C.~A. Parkinson, F.~Markowetz, and J.~D. Brenton
  (2015).
\newblock Spatial and temporal heterogeneity in high-grade serous ovarian
  cancer: A phylogenetic analysis.
\newblock {\em PLoS Med\/}~{\em 12\/}(2), e1001789.

\bibitem[\protect\citeauthoryear{Tehrani}{Tehrani}{2013}]{Tehrani2013}
Tehrani, J.~J. (2013).
\newblock The phylogeny of little red riding hood.
\newblock {\em PLoS ONE\/}~{\em 8\/}(11), e78871.

\end{thebibliography}
